\documentclass[conference]{IEEEtran}
\IEEEoverridecommandlockouts
\usepackage{cite}
\usepackage{amsmath,amssymb,amsfonts}
\usepackage{algorithmic}
\usepackage{graphicx}
\usepackage{textcomp}
\usepackage{xcolor}
\usepackage{outline}
\usepackage{booktabs}
\usepackage{subcaption}
\usepackage{listings}
\usepackage{bbding}
\usepackage{pifont}
\usepackage{wasysym}
\usepackage{amssymb}

\usepackage{array} 

\def\BibTeX{{\rm B\kern-.05em{\sc i\kern-.025em b}\kern-.08em
    T\kern-.1667em\lower.7ex\hbox{E}\kern-.125emX}}
\begin{document}
\title{RFRL Gym: A Reinforcement Learning Testbed for Cognitive Radio Applications}

\author{\IEEEauthorblockN{Daniel Rosen\IEEEauthorrefmark{1}, Illa Rochez\IEEEauthorrefmark{1}, Caleb McIrvin\IEEEauthorrefmark{1}, Joshua Lee\IEEEauthorrefmark{1}, Kevin D’Alessandro\IEEEauthorrefmark{1}, Max Wiecek\IEEEauthorrefmark{1}, Nhan Hoang\IEEEauthorrefmark{1},\\
Ramzy Saffarini\IEEEauthorrefmark{1}, Sam Philips\IEEEauthorrefmark{1}, Vanessa Jones\IEEEauthorrefmark{1}, Will Ivey\IEEEauthorrefmark{1}, Zavier Harris-Smart\IEEEauthorrefmark{2}, \\
Zavion Harris-Smart\IEEEauthorrefmark{2}, Zayden Chin\IEEEauthorrefmark{2}, Amos Johnson\IEEEauthorrefmark{2}, Alyse M. Jones\IEEEauthorrefmark{1}, William C. Headley\IEEEauthorrefmark{1}}
\IEEEauthorblockA{\IEEEauthorrefmark{1}Virginia Tech National Security Institute, \IEEEauthorrefmark{2}Morehouse College \\
amos.johnson@morehouse.edu, alysemjones@vt.edu, cheadley@vt.edu}}

\maketitle

\begin{abstract}
Radio Frequency Reinforcement Learning (RFRL) is anticipated to be a widely applicable technology in the next generation of wireless communication systems, particularly 6G and next-gen military communications. Given this, our research is focused on developing a tool to promote the development of RFRL techniques that leverage spectrum sensing. In particular, the tool was designed to address two cognitive radio applications, specifically dynamic spectrum access and jamming.  In order to train and test reinforcement learning (RL) algorithms for these applications, a simulation environment is necessary to simulate the conditions that an agent will encounter within the Radio Frequency (RF) spectrum. In this paper, such an environment has been developed, herein referred to as the RFRL Gym. Through the RFRL Gym, users can design their own scenarios to model what an RL agent may encounter within the RF spectrum as well as experiment with different spectrum sensing techniques. Additionally, the RFRL Gym is a subclass of OpenAI gym, enabling the use of third-party ML/RL Libraries.

We plan to open-source this codebase to enable other researchers to utilize the RFRL Gym to test their own scenarios and RL algorithms, ultimately leading to the advancement of RL research in the wireless communications domain. This paper describes in further detail the components of the Gym, results from example scenarios, and plans for future additions.
\end{abstract}

\begin{IEEEkeywords}
machine learning, reinforcement learning, wireless communications, dynamic spectrum access, OpenAI gym
\end{IEEEkeywords}

\section{Introduction}

Due to the rapid growth of Radio Frequency (RF) spectrum usage, efficient use of the spectrum has become increasingly difficult, leading to over-congestion in the RF spectrum \cite{spectrum_management}. This largely is a consequence of the exponential increase in wireless devices, with the CTIA citing a growth of 183 times its size since 2010 in its 2021 Wireless Internet Survey~\cite{ctiaWirelessSurvey}, ultimately resulting in decreased data rates from poor channel conditions and collisions \cite{api}. These collisions are a form of signal jamming, also known as interference, leading to incomplete and corrupted transmissions. Interference can either be intentional from an adversary or unintentional due to less-than-ideal conditions in the RF spectrum. Furthermore, spectrum congestion makes it more difficult to differentiate whether the collision was from an adversary or over-congestion. Regardless, to avoid further loss to communication systems, both interference types must be prevented. Thus, to prevent interference as well as losses in data rate, there is a clear need for improvement in the utilization of the RF Spectrum. 

In recent years, Cognitive Radios (CR) have gained traction as a promising solution to alleviating spectrum congestion \cite{Mitola} due to its ability to intelligently determine which channels are currently occupied, and which are not through Dynamic Spectrum Access (DSA)\cite{defofCR}. DSA was proposed to overcome the shortcomings of traditional methods, such as Frequency Hopping Spread Spectrum (FHSS)\cite{Khatib_2011} and Direct Sequence Spread Spectrum (DSSS)\cite{dsss}, where both fail in a dynamic RF spectrum from not adapting its method to where holes in the spectrum are located. DSA, however, is able to do so through spectrum sensing, where methods such as energy detection are utilized to identify which frequency bands are in use \cite{cycle}, allowing a CR to switch to an empty band. However, DSA alone is not enough to handle spectrum congestion since it cannot \textit{predict} the location of future spectrum holes, resulting still in accidental collisions. However, by coupling DSA with predictive intelligence, a CR can predict where and when other users will occupy the RF spectrum in the future. One promising solution is Reinforcement Learning (RL).

RL is a Machine Learning method known for its ability to predict the best decisions to achieve a desired outcome\cite{RLbook}. When applied to CRs, RL's predictive capabilities enable signals to preemptively respond to the movements of other signals in the spectrum as well as changing channel conditions such that the optimal frequency at which to transmit can be identified. Significant effort has been invested into investigating RL-based protocols for spectrum assignment and wireless communication enhancement, where RL has shown to drastically reduce congestion in the spectrum \cite{2012_jamming,COPSQ,2016_wacr,DRLInWireless}. 

From these works, it is clear that RL Algorithms \emph{must} be trained in a representative environment, in this case the RF spectrum. However, there is currently an absence of a comprehensive, universal tool simulating the RF spectrum that researchers can use to develop and test RL algorithms for CR-specific applications. Conventionally, new tools have been developed for each project and can not be easily used in other applications. Thus, this paper introduces the RFRL Gym\footnote{https://github.com/vtnsiSDD/rfrl-gym}, a training environment of the RF spectrum designed to provide comprehensive functionality, such as custom scenario generation, multiple learning settings, and compatibility with third-party RL packages. As a result, the RFRL Gym can be applicable to a wide range of wireless communications projects, where researchers can utilize the tool to further the advancement of RL in wireless communications. Additionally, through a gamified mode of the RF spectrum, this tool can be used to teach novices about the fields of AI/ML and RF. The remainder of this paper defines the current state of the RFRL Gym, its capabilities, and planned future enhancements.
\section{Background}
As mentioned, prior works \cite{gr_gym,grgym_ieee,DRLMRINETGYM,ns3_gym} have developed tools to achieve RL for their own wireless communications applications. Table \ref{tab:features} describes how the RFRL Gym differs from these existing tools through the following notable key features. 

\textbf{Flexible Scenario Design:} The ability to create, modify, or utilize different scenarios to view how RL algorithms will react and interact in different learning spaces. \cite{DRLMRINETGYM} was designed specifically for a particular use case of wireless communications and contained no ability to create custom scenarios on their framework. \cite{ns3_gym} did have some included scenarios within their gym, such as for WiFi and cellular scenarios. However, it is unclear if custom scenarios can be generated and if so, how difficult it would be. \cite{gr_gym,grgym_ieee} contain the most flexibility in scenario generation of the three since its data is generated through GNU Radio, a platform to design communication systems \cite{gnuradio}. However, this process is not straightforward, as designing flow-graphs in GNU Radio can be difficult for non-experts and those not familiar with the software. Therefore, the RFRL Gym includes flexible scenario generation where users can easily create custom scenarios according to their application through input json files.

\textbf{RL Package Compatibility: }The ability to interface with other RL libraries to expand a user's access to different RL agents. Examples include Mushroom RL\cite{JMLR:v22:18-056} and Stable Baselines\cite{stable-baselines}. This feature removes the need to implement RL algorithms manually if not desired. Of the tools considered, only the RFRL Gym has explicitly tested and confirmed this feature. Additionally, custom creation of RL algorithms is also possible within the RFRL Gym.




\textbf{Spectrum Sensing Capabilities:} The ability to locate spectrum holes through spectrum sensing. To fully realize a DSA scenario, this is a requirement. \cite{ns3_gym} does have capabilities for spectrum sensing, but to our knowledge, there is no ability to create custom spectrum sensing methods.

\textbf{Multi-Agent Capabilities:} The ability to support several intelligent agents (i.e. RL agents). \cite{DRLMRINETGYM} and \cite{ns3_gym} do have capabilities for multi-agent reinforcement learning. It is our desire for future work to include this functionality within the RFRL Gym.

\textbf{Ease Of Use:} The tool is designed with considerations for an easy user experience, especially for those with little or no familiarity with wireless communications and/or RL. This feature allows for ML/RL specialists to better interact and interface with Wireless Communication specialists through easy scenario generation, provided examples, and a gamified mode to learn the wireless communications concepts at an abstracted level. By doing so, the RFRL Gym suits the needs for both novices to learn RFRL concepts and experts to use the gym to further their research.

\textbf{Graphical User Interface:} The ability to control the simulation (i.e. scenario generation) through a Graphical User Interface (GUI) to make it easier for users to interface with the RFRL Gym. Existing tools are all controlled through scripts and a command line interface. If the user wants to make scenario generation changes, they are forced to change it within the code. However, a GUI bypasses this need such that users can make any changes to the scenario without touching the code. Future plans for the GUI include controlling RL hyperparemeters and environment parameters such that the entire simulation can be run entirely from the GUI.

\textbf{Hardware Compatible:} The ability to operate over radio hardware in real-time, rather than simulation alone, for a realistic implementation. \cite{gr_gym,grgym_ieee} are hardware compatible since GNU Radio supports radio hardware. Additionally, a primary contribution of \cite{DRLMRINETGYM} was to implement their system over embedded systems in radio hardware. In future versions, the RFRL Gym will also aim for hardware compatibility. 

In comparison to existing tools described above, the RFRL Gym was designed to serve as a universal, easy-to-use tool that both novices and experts within this field can use to learn about AI/ML and RF as well as further technological advancement of RFRL. For novices, in the gamified mode, the complex features of wireless communications within the RFRL Gym are abstracted so they can use the tool with limited knowledge of wireless communications and be eased into more advanced topics, such as signal modulation and the generation of IQ data. In addition, the gym has a built-in GUI system for setting scenarios and parameters without changing the code as well as the ability to support outside packages, such as Mushroom RL. Additionally, prior works discussed in this section only allow for pre-set scenarios, whereas our tool allows for scenario customization. These additions separate our tool from others, emphasizing an easy-to-use environment for experimentation and testing.

  \begin{table}[t]
\caption{Comparison of existing OpenAI Gym RL tools for wireless communications}
\begin{tabular}{p{3.5cm} p{0.6cm} p{0.9cm} p{1.1cm} p{0.5cm} p{1.0cm}}
\toprule
\vspace{4mm}\textbf{Features} & 
\textbf{RFRL Gym} & \textbf{GrGym} \cite{gr_gym}, \cite{grgym_ieee} & \textbf{MR-iNet Gym} \cite{DRLMRINETGYM} & \textbf{ns3-Gym} \cite{ns3_gym}\\ \midrule

Flexible Scenario Design & \hspace{1mm} \Checkmark & \hspace{1mm} \Checkmark & \vspace{0.5mm} & \hspace{1mm} \Checkmark\\
\vspace{1mm}
RL Package Compatibility & \vspace{1mm} \hspace{1mm} \Checkmark\\

\vspace{1mm}
Spectrum Sensing Capabilities & \vspace{1mm} \hspace{1mm} \Checkmark & \vspace{1mm} & \vspace{1mm} & \vspace{1mm} \hspace{1mm} \Checkmark\\

\vspace{1mm}
Multi-Agent Capabilities & \vspace{1.0mm} & \vspace{1mm} & \vspace{1mm} \hspace{2mm} \Checkmark & \vspace{1mm} \hspace{1mm} \Checkmark\\

\vspace{1mm}
Ease of Use & \vspace{1mm} \hspace{1mm} \Checkmark & \vspace{1mm} & \vspace{1mm} & \vspace{1mm}\\

\vspace{1mm}
Graphical User Interface & \vspace{1mm} \hspace{1mm} \Checkmark & \vspace{1mm} & \vspace{1mm} & \vspace{1mm}\\

\vspace{1mm}
Hardware Compatible & \vspace{1mm} & \vspace{1mm} \hspace{1mm} \Checkmark & \vspace{1mm} \hspace{2mm} \Checkmark & \vspace{1mm} \hspace{1mm}\\\bottomrule
\end{tabular}
\label{tab:features}
\end{table}

\begin{figure*}[h!tbp]
    \centering
    \includegraphics[scale = 0.5]{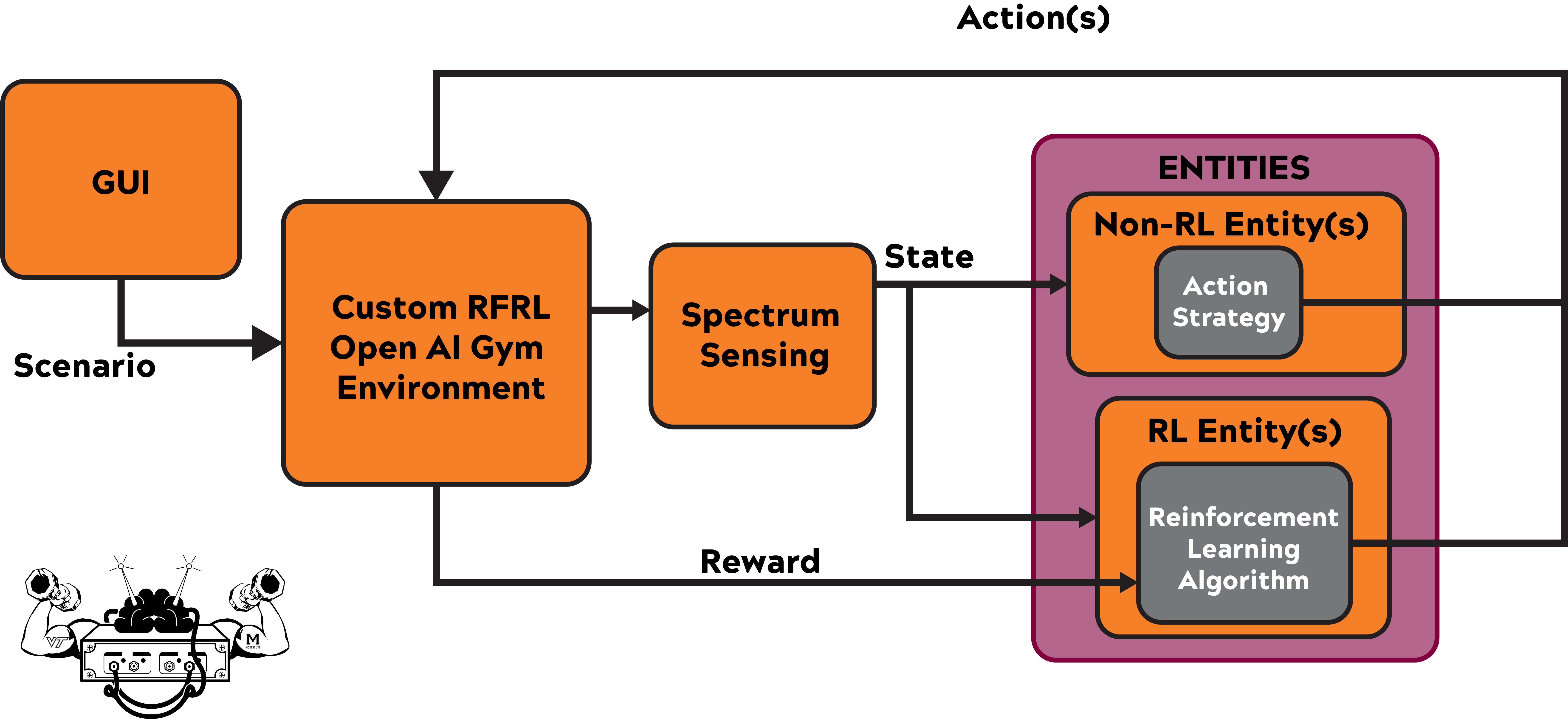}
    \caption{The RFRL Gym framework, illustrating how user-defined scenarios and reinforcement learning algorithms interact with the gym environment. This framework facilitates simulating an RF-based reinforcement learning interaction loop of the RL agent taking in states (e.g. sensing results) and rewards (e.g. receiver performance) and providing to the gym its determined next action (e.g. transmitter tuning).}
    \label{fig:system_diagram}
\end{figure*}

\section{The RFRL Gym Framework}

A high-level overview of the structure of the RFRL Gym is provided by Figure \ref{fig:system_diagram}. The framework consists of a series of interconnected modules that function together in order to simulate RF spectrum scenarios. These modules follow the Open AI Gym API to ensure the RFRL Gym is compatible with third party gym-compatible RL software through its predefined functions and methods for a Reinforcement Learning agent and the environment to communicate. As a result, the RFRL Gym is able to exploit the ``RL Cycle" training protocol. This protocol is a cyclical passage of information between the Gym and RL agent over a discrete set of training episodes, with each episode consisting of a number of time-steps.  The cycle begins at the beginning of each time-step when the RL agent performs an action in the custom-made RF environment.  This action may be to occupy a channel or to not transmit in the next timestep.  At this point, Non-RL Entities will also select an action.  Based on the actions of the Non-RL Entities, the RL Agent's action is evaluated and assigned a reward.  This reward and the new observation space are made accessible to the RL Agent for training.  This process repeats for all time-steps. The remainder of this section describes the individual components of the RFRL Gym outlined in Fig. \ref{fig:system_diagram}.

\subsection{Non-Player Entities}
Several non-player entities are implemented in our environment. Non-Player refers to an entity whose decision-making is not controlled by the primary Reinforcement Learning algorithm.  They will herein be referred to as ``Entities".  These Entities are crucial for testing as they simulate physical interactions between signals and replicate real signals in the environment. There are five entities that have been implemented into our code base at this point, as described in Table \ref{tab:entities}.

\begin{table}[t]
\caption{Non-player entity types within the RFRL Gym environment that can be included in scenario files, with descriptions and real-world use cases}
\begin{tabular}{@{}lp{3.3cm}ccc@{}}
\toprule
\textbf{Entity Type} & \textbf{Description} & \textbf{Real-world Example}                          \\ \midrule
\vspace{1mm}
Constant       & Remains in one channel & T.V. or Radio                           \\
\vspace{1mm}

Fixed Hopper   & Moves according to a user-defined pattern & Bluetooth          \\
\vspace{1mm}
Stochastic Hopper   & Channel selected based on a set of user-defined probabilities for each channel & N/A          \\
\vspace{1mm}
Agile Hopper & Moves to an empty channel. Has perfect knowledge of other entities in the bandwidth. If multiple channels are empty, it selects one with uniform probability. Most similar to the RL agent but without the RL. & CR with DSA \\
\vspace{1mm}
Simple Jammer & Has perfect knowledge of other entities in the environment, and moves to an occupied channel based on uniform distribution.  If no channels are occupied, it moves to an unoccupied channel. & Intelligent Adversary \\
Custom & Created by users of our codebase & N/A 
\\\bottomrule
\end{tabular}
\label{tab:entities}
\end{table}

\subsection{RL Entities}

The purpose of the RFRL Gym is to train RL agents for RF applications.  Users can run simulations with their own custom RL algorithms or those built by third parties. Multiple modes have been implemented for the observation and reward to enable a diversity of scenarios on which to train RL agents.   


\subsubsection{Reward Modes} The two reward types currently implemented in the environment for the RL agent are DSA reward and Jamming reward. DSA mode can be used to simulate a situation where an RL agent wants to avoid interference, either by avoiding a channel with bad channel conditions or a channel with other users and/or adversaries. Applications include commercial communications as well as Electronic Warfare (EW). Alternatively, Jamming mode allows for the simulation of situations where the agent wants to prevent others from successful communications and use of the RF spectrum by acting as an adversary. This mode is also applicable to EW. The optional dampening factor was added to consider the computational cost of switching to model the desired real-world behavior of energy conservation, where the ideal behavior is to move only when necessary. This is especially important for small, energy-constrained devices, such as IoT devices.        \begin{itemize}
        \item DSA Reward - Rewards the agent for avoiding channels containing other entities. 
        \begin{equation}
        \textbf{DSA reward } =
            \begin{cases}
                1, &  \text{no collision} \\
                -1, &  \text{collision}
            \end{cases}
        \end{equation}
        
        \item Jamming Reward - Rewards the agent for occupying the same channel as a particular entity in the environment. The entity to be jammed is selected by the user.
        \begin{equation}
        \textbf{Jamming reward = }
            \begin{cases}
                1, &  \text{collision} \\
                -1, &  \text{no collision}
            \end{cases}
        \end{equation}
        \item Dampening Factor - The dampening factor will increase the reward if the agent stays in the same channel and will reduce the reward if it changes channels.
        \begin{equation}
        \textbf{reward} =
            \begin{cases}
                reward - 0.5, &  \text{movement made}\\
                reward + 0.5, &  \text{no movement made}
            \end{cases}
        \end{equation}
        \end{itemize}

        \begin{itemize}
        \item \textbf{DSA Reward} - Rewards the agent for avoiding channels containing other entities. The agent is awarded a $1$ for no collision and $-1$ for a collision.
        
        \item \textbf{Jamming Reward} - Rewards the agent for occupying the same channel as a target entity in the environment. The entity to be jammed is selected by the user. The agent is awarded a $1$ for a collision and $-1$ for no collision.
        \item \textbf{Dampening Factor} - The dampening factor will increase the reward if the agent stays in the same channel and reduce the reward if it changes channels. The agent is awarded an additional $0.5$ for no movement made and $-0.5$ for movement made.
        \end{itemize}

\subsubsection{Observation Modes} The gym contains two different observation modes for the agents: detect and classify. These modes determine what information is sent to the agent as the current state of the environment. In detect mode, the agent only knows if a channel is occupied or not. In this mode, the RL agent does not know what type of entities or how many entities occupy a single channel. In classify mode, the agent receives information about the channel location, type, and number of entities (i.e. which entities are in which channel or if there is a Multi-Entity collision in a channel). These modes are realized through spectrum sensing, where occupancy can be determined across each frequency based on the amount of energy present. If the amount of energy is above the noise floor, it is an indication that an entity is transmitting over that channel and/or the channel has poor conditions.

\subsection{Rendering Modes}
The RFRL Gym provides two rendering modes to visualize simulations, the state of entities and their interactions at each time-step, as well as results: terminal rendering mode and PyQt rendering mode. Both modes display the location of the entities within the different channels across time as well as the reward for each time step and cumulative reward across each episode.

\subsubsection {Terminal Rendering}
The terminal rendering mode sends data to be displayed in Standard Output using ascii text. Therefore, the data can be directed to the terminal or redirected to a file. Depending on whether the gym is in classify or detect mode, the visualization will change. In detect mode, occupied channels are represented with a 1, while unoccupied channels are represented with a 0. In classify mode, the location of each entity is shown with its corresponding entity number. Lastly, the terminal mode runs efficiently, making it suitable for more complex scenarios where a faster output is desired. 
    
\subsubsection {PyQt Rendering}
Enabled by the \textit{pyqtgraph} package, the PyQt rendering mode displays the same data as the terminal mode but with more detailed visuals, making it easier for users to visually understand how entities move across time and frequency within the RF spectrum. This mode also displays a plot of the cumulative reward for each time-step in an episode. An example of visual results for this mode can be seen in Section IV. 
    
\subsection{Scenarios}
A scenario is the set of parameters that define the environment in which the agent is placed to perform its learning. The parameters fall into three categories: \emph{environment}, \emph{render}, and \emph{entities}.  Scenarios are defined in customizable JSON files ("scenario files") that are fed into the RFRL Gym.
\subsubsection{Environment} This section defines the learning parameters needed to prepare the environment, such as the number of channels available and number of time steps. Additionally, the observation mode can be set to the ``classify" or ``detect" mode, and similarly, the reward mode can be set to ``dsa" for dynamic spectrum access, or ``jam" for jamming mode. For jamming mode, the target entity is also set.

\subsubsection{Entities} Here, the user will determine which entities will exist in the environment. An entity is designated by a user-defined name, which will appear in the rendered output. Within an entity, various parameters can be set. For the gamified, basic mode, these parameters include the channel(s) that the entity can occupy, the on/off transmission pattern(``on" referring to a non-idle state, ``off" referring to idle), and the first and last steps at which the entity transmits. For the advanced mode containing IQ, the entities' modulation type and order, center frequency, bandwidth, start time, and duration can also be set. 
       
\begin{figure*}[t]
    \centering
    \includegraphics[width=0.95\textwidth]{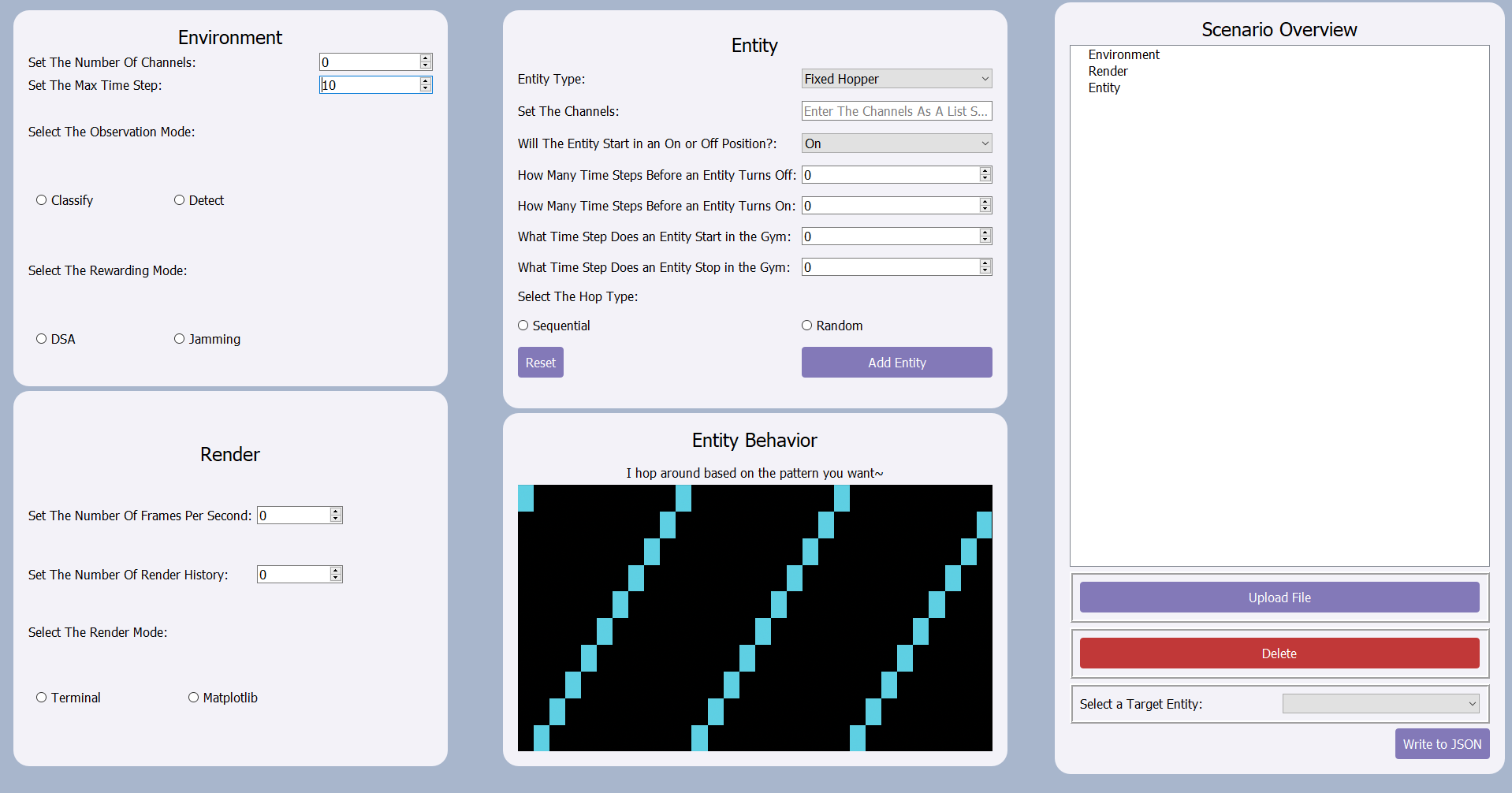}
    \caption{GUI-based scenario generator where users can set up the environment, render, and entities.}
    \label{fig:gui}
\end{figure*}

\subsubsection{Render} This section determines whether the render is displayed within the terminal, or if it utilizes the render written using the \textit{pyqtgraph} library. The render's speed, in frames per second, and the number of steps that the render displays at a time are also included options.

To improve the user experience, a Scenario Selection GUI was created which enables the modification and creation of scenario files without interfacing with a JSON file directly. This simplifies the process of creating scenario files and allows non-software-oriented users to produce scenarios.  Fig. \ref{fig:gui} shows the layout of the GUI.

\section{Simulation Results and Discussion}
In this section, multiple representative wireless communications scenarios are presented in order to highlight the capabilities of the RFRL Gym for test and evaluation of RL agents. For all scenarios, the environment consists of 10 wireless channels, assumes learning episodes consisting of 100 time-steps, 100 training episodes, a Q-learning agent balancing exploration and exploitation through an epsilon greedy strategy, and the use of the MushroomRL software package for training and evaluating the agent.

\begin{table}[hbt!]
\caption{Scenario settings with entities, reward, dampening, and observation modes defined}
\begin{tabular}{@{}lp{1.5cm}ccccc@{}}
\toprule
\textbf{Scenario} & \textbf{Entities} & \textbf{Reward} & \textbf{Dampening} & \textbf{Observation}    \\ \midrule
\vspace{1mm}
\textbf{1} & 1 Constant & Jamming & No & Classify \\
\vspace{1mm}
\textbf{2} & 1 Fixed Hop & Jamming & No & Classify\\ 
\textbf{3} & 1 Fixed Hop  & DSA & No & Detect\\ 
& 1 Constant &&&\\
\vspace{1mm}
& 1 Stochastic &&&\\
\textbf{4} & 1 Fixed Hop  & DSA & No & Classify\\ 
& 1 Constant &&&\\
& 1 Stochastic &&&\\
\vspace{1mm}
& 1 Agile &&&\\
\bottomrule
\end{tabular}
\label{tab:scenarios}
\end{table}

\subsection{Scenario 1: Single Entity Jamming}
\begin{figure}[hbt!]
\centering
    \includegraphics[scale = 0.4]{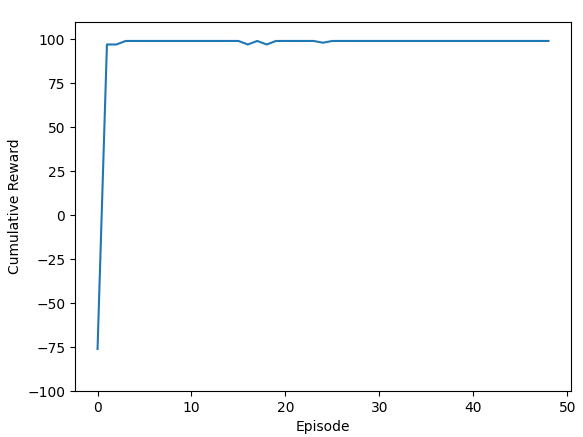}
    \caption{Cumulative reward results for Scenario 1, showing optimal behavior by converging to the optimal reward of 100}
    \label{fig:scenario_1_results}
\end{figure}

This is a simple scenario that was designed to be solved easily and validate that the RFRL Gym performs as expected and yields accurate results. It consists of a single Constant Entity that is to be jammed.  As expected, a Q-Learning agent was able to converge to an optimal policy quickly within the first five episodes, as demonstrated in Fig. \ref{fig:scenario_1_results}.

\subsection{Scenario 2: Hop Frequency Jamming|Non-Markovian}
\par This scenario consists of a single Hopping entity with a fixed hopping pattern.  It spends 2 time-steps in channel 4, 1 time-step in channel 5, and repeats.  The object of this scenario is to produce a \emph{Non-Markovian} problem, meaning the problem does not follow the Markov Decision Process which holds the assumption of the Markov property where the evolution of the Markov process depends only on the current state. In this case, the current state is not always sufficient information for the next time step, hence the agent cannot achieve an optimal policy.   In order for the agent to jam the Hopper perfectly, it must have some sense of the previous location of the hopper. For example, if the Hopper is in channel 4, the agent would need to know whether the Hopper was in channel 4 or 5 in the previous time-step.  Without that information, our Q-Learning agent can not predict whether the entity will stay or move.  Therefore, it will converge to a sub-optimal policy.

\par
\begin{figure}[hbt!]
\centering
    \includegraphics[scale = 0.4]{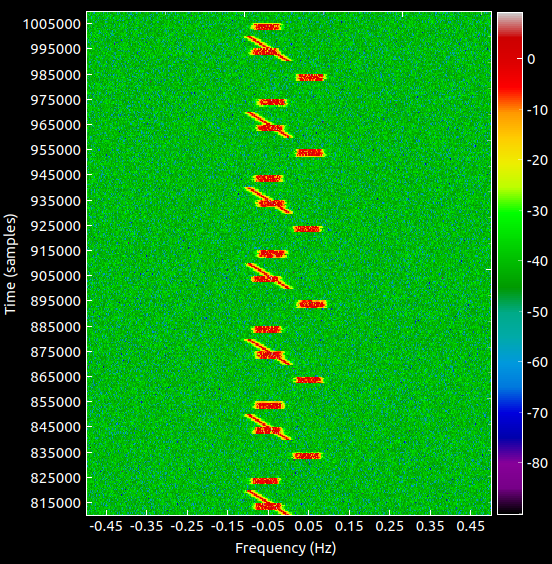}
    \caption{PyQt render results for scenario 2, shown in advanced mode where IQ data is viewed. The diagonal signal is the RL agent, while the horizontal signal is the non-RL entity the agent is targeting to jam.}
    \label{fig:scenario_2_render}
\end{figure}

 Optimally, the agent would jam the Hopper in all 100 time-steps. In Fig. \ref{fig:scenario_2_render}, the agent can be seen choosing not to transmit in the second and third time-steps of the hopping pattern.  This is because it was unable to consistently jam the Hopper  during the training process due to Non-Markovian behavior.  These incorrect guesses yield a reward of $-1.0$ but non-transmission yields a reward of $0.0$.  Thus, the agent elects to accept the suboptimal reward of 33 and not risk the negative reward.  Calculations below show this suboptimal convergence to 33, rather than the optimal performance of 100.
\[\textbf{Sub-Optimal Reward: } 100\Big(\frac{1}{3}(1) + \frac{2}{3}(0)\Big)= 33\]

\par The results produced by the RFRL Gym were aligned exactly with the expected outcome.  This scenario verifies that our Gym will perform as expected even when an agent is unable to achieve an optimal policy.  The dependability of the Gym's results make it a valuable tool for formulating and testing hypotheses.

\subsection{Scenario 3: Solvable Multi-Entity DSA}
\par
\begin{figure}[hbt!]%
\centering
\begin{subfigure}{\columnwidth}
\includegraphics[width=\columnwidth]{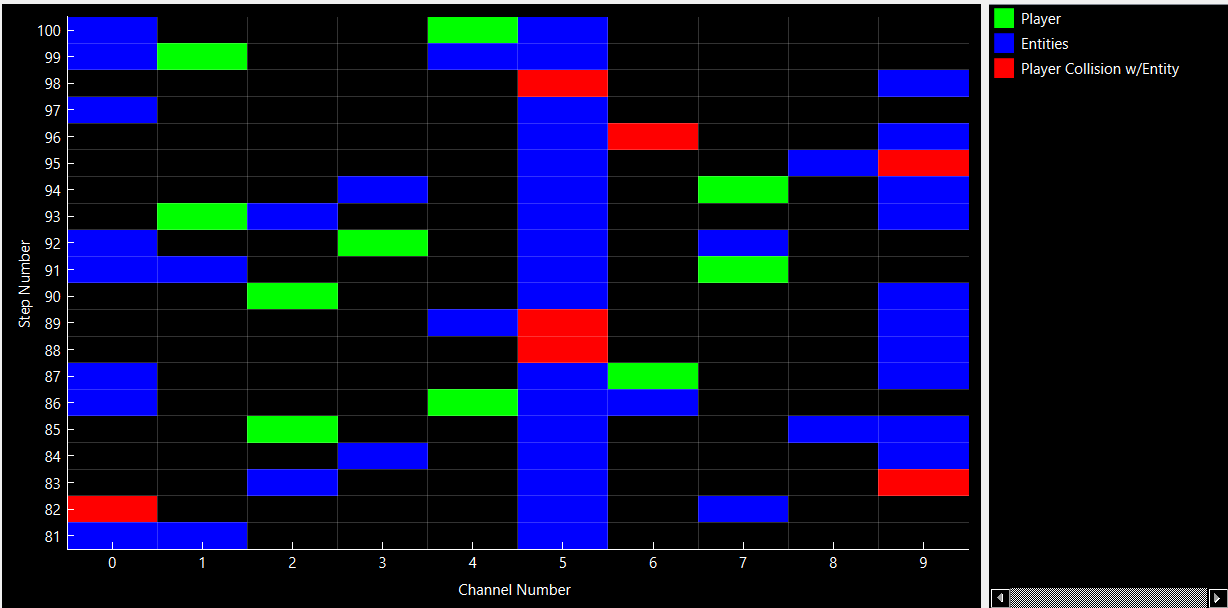}%
\caption{Render results showing non-convergent behavior before learning, where the RL agent collides with other entities within the environment\hfill \break}%
\label{fig:before}%
\end{subfigure}
\\
\begin{subfigure}{\columnwidth}
\includegraphics[width=\columnwidth]{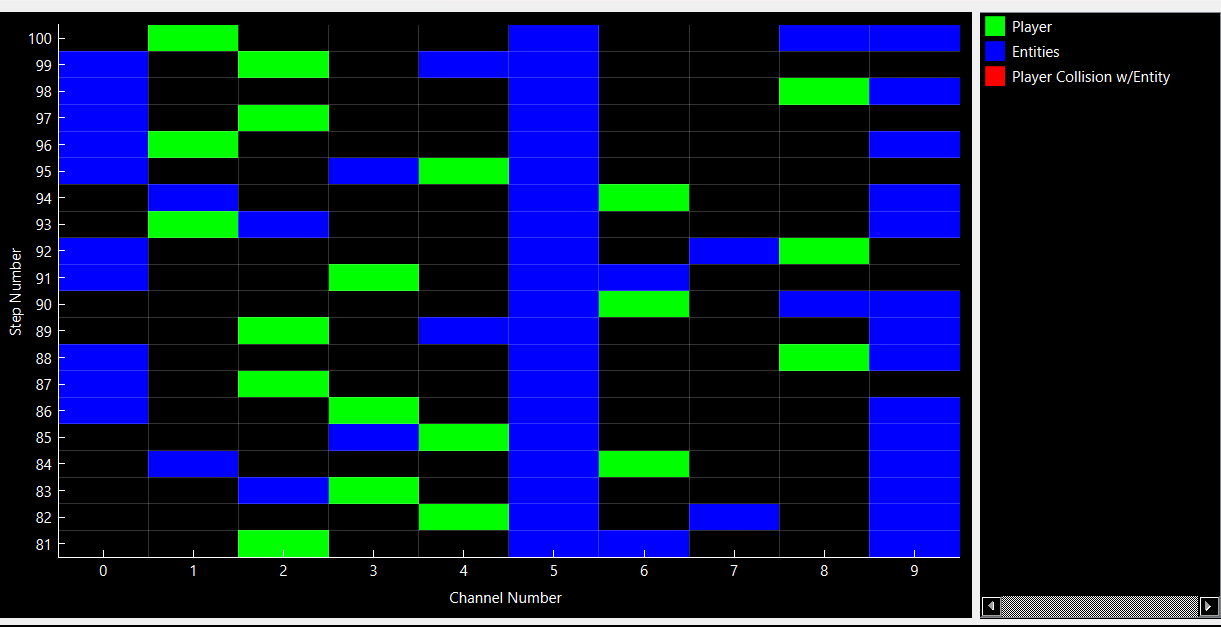}%
\caption{Render results showing convergent behavior after learning, where the RL agent successfully avoids other entities within the environment}%
\label{fig:after}%
\end{subfigure}\hfill%
\caption{Scenario 3 render results, showing before and after convergence behavior, in the gamified detect mode.}
\label{fig:scenario_3_render}
\end{figure}

   This scenario contains a Fixed Hopper Entity moving through all channels non-sequentially, a Constant Entity in channel 5, and a Stochastic Hopper that moves to either channel $0$ or $9$ with equal likelihood. This scenario exhibits the multi-Entity capability of our environment.  As shown in Fig. \ref{fig:scenario_3_render}, the Q-Learning agent can handle avoidance of multiple Entities and achieve an optimal policy. The RFRL Gym's gamified rendering functionality enables easy visualization of the agent's optimal policy. The absence of collisions after convergence can be seen easily by the lack of red blocks in Fig. \ref{fig:after}, as opposed to before convergence where collisions occurred frequently in Fig. \ref{fig:before}.  

A notable conclusion that can be drawn from this result as well is that a Q-Learning agent in DSA mode is able to achieve an optimal policy when entities behave according to a stationary distribution, meaning the probability in which it chooses each channel stays constant.  However, this is not true for Agile Entities as will be shown in Scenario 4.

\subsection{Scenario 4: Unsolvable Multi-Entity DSA}

\begin{figure}
\includegraphics[width=\columnwidth]{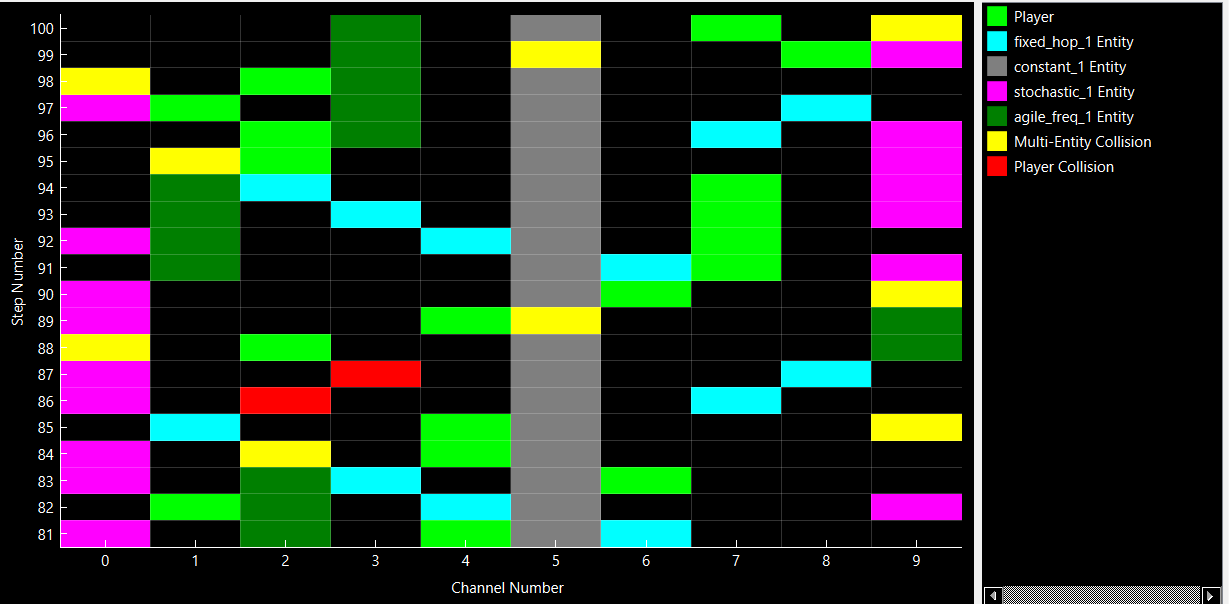}
\caption{Render results showing non-convergent behavior due to collisions with an Agile Entity.  Render is in classification mode to show that the Agent is colliding with the Agile Entity.}%
\label{fig:scenario_4_render}%
\end{figure}

\begin{figure}[t]
\centering
    \includegraphics[scale = 0.55]{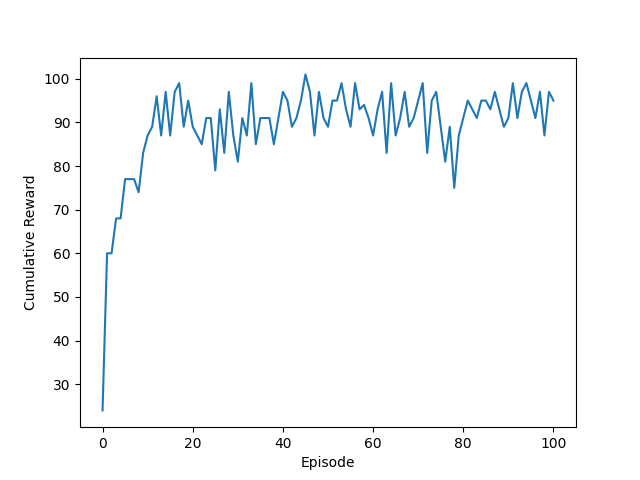}
    \caption{Cumulative reward results for Scenario 4, showing suboptimal behavior due to the Agile Entity}
    \label{fig:scenario_4_results}
\end{figure}

This scenario is similar to Scenario 3.  The single difference is the addition of an Agile Entity moving across all channels. The Stochastic Hopper Entity still behaves according to a stationary distribution between each of its channels since it only moves between the two channels with the highest probabilities (channel 0 and 9), which the Q-learning agent learns to avoid. On the other hand, 
the addition of the Agile Entity confuses the RL agent such that it is unable to definitively predict where it will go next because the Agile Entity has no deterministic pattern due to its non-stationary, time-varying behavior caused by randomly choosing an empty channel according to where other entities occupy other channels. Therefore, the RL agent will occasionally collide with the Agile Entity, as can be seen in red in Fig. \ref{fig:scenario_4_render}. As a result, the agent is unable to converge to an optimal policy, as can be seen by the noisy, non-stabilizing results in Fig. \ref{fig:scenario_4_results}. This result implies that more advanced Reinforcement Learning algorithms such as Deep Q Learning are necessary to solve problems with non-stationary, time-varying behavior.

\hfill%

\section{Future Additions}

\subsection{Multi-Agent RL functionality}
In our gym, we are currently moving towards incorporating multi-agent functionality through a separate, multi-agent-specific environment.  As realistic dynamic spectrum access scenarios may involve highly contested bandwidths, it is important to consider and provide solutions for cases where several RL agents interact in the same set of channels. As a result, incorporating Multi-Agent Reinforcement Learning (MARL) learning strategies is crucial to the success of our agents in realistic scenarios. These scenarios may be divided into two separate classes - cooperative and adversarial.

\subsubsection{Cooperative}Cooperative MARL agents attempt to coordinate with other agents in order to maximize total uncontested access to the spectrum.
\subsubsection{Adversarial} Adversarial agents attempt to prevent other agents from accessing the spectrum by intentionally broadcasting on the channel already occupied by its target or one the target is predicted to hop to in the future. 

Incorporating MARL into the RFRL Gym could be made possible by utilizing MARL libraries compatible with OpenAI Gym or similar to OpenAI Gym. One such package worth investigating is PettingZoo \cite{pettingzoo}, a MARL library designed with OpenAI Gym's simple API in mind in addition to the ability to create custom MARL environments.

\subsection{Physical signal integration and Hardware compatibility}

In real-world scenarios, perfect knowledge of the environment is unlikely. Thus, the RFRL Gym will include functionality for converting raw signal data into state information that can be processed by RL agents. Convolutional neural networks (CNNs) are currently being developed to identify signal types based on a signal's structure \cite{cnn}. The network will take in raw waveform IQ data from an RF channel and classify the signal. This result can be integrated into the observation space of RL agents to target specific entities. Additionally, we are working towards developing the hardware functionality for more realistic software simulation such that signals experience realistic channels conditions in which fading and multi path propagation exist.

\subsection{GUI expansion}

The next iteration of the GUI intends to connect the visualizations rendered with PyQt to the pre-existing Scenario Selection GUI so that the user can run simulations entirely from the Scenario Selection GUI. Additionally, we intend to implement tabs that will guide the user through the scenario setup process, rather than having them interact with all of the steps simultaneously. Also, there are plans to implement an auto-sync feature. This would mean that whenever a new rewarding type, observation mode, or non-player entity is added, the visual output will automatically update to allow the user to see their changes. 





\section{Conclusion}

A sharp rise in wireless usage has caused significant congestion in the RF spectrum. To remedy this, the predictive capabilities of Reinforcement Learning are increasingly being exploited in Cognitive Radio applications.  Therefore, the existence of a resource like the RFRL Gym is imperative to train algorithms in a representative environment.
 
Tools that serve similar purposes have been created, but the RFRL Gym possesses various distinct advantages and includes more comprehensive functionality.  Firstly, the simulation functionality in our environment is abstracted, so technical expertise in Wireless Communications is not necessary to use the Gym. Secondly, our codebase is compatible with external libraries including Mushroom RL and Stable Baselines.  A third major advantage of our Gym over comparable resources is the GUI which makes the operation of the environment simple. Though the RFRL Gym currently possesses extensive functionality, additions are still being made.  Future iterations of the environment will show MARL functionality, physical signal integration, radio hardware compatibility, and simulations run entirely from the GUI. 

All in all, we hope our work will aid in developing new reinforcement learning strategies for cognitive radio applications. By providing a powerful tool with a low barrier of entry, we seek to encourage the exploration of new pursuits in the rapidly growing field of RFRL.
\bibliographystyle{IEEEtran}
\IEEEtriggeratref{12}
\bibliography{bibliography.bib}
\end{document}